\documentclass{article}%
\usepackage{amsmath}
\usepackage{amssymb}
\usepackage{amsfonts}
\usepackage{graphicx}%
\providecommand{\U}[1]{\protect\rule{.1in}{.1in}}

\newlength{\defaultparindent}
\newenvironment{Body Text}{}{}

\newenvironment{Default Paragraph Font}{}{}
\begin{document}

\title{The study of temperature field impact on velocity of fluid in streamlines coordinates in free convection problem }
\author{Sergiej Leble* and Witold M.Lewandowski,\\*Immanuel Kant Baltic Federal University*,\\
236041, ul. Al. Nevskogo, Kaliningrad, Russia,\\Gda\'{n}sk University of Technology}
\maketitle

\begin{abstract}
A transition to coordinates formed by streamlines and orthogonal ones are convenient to simplify the Navier-Stokes and Fourier-Kirchoff system. We derive transformation to such coordinates, taking into account a necessity to introduce integrating factor that is not equal to unity in a viscous flow, The transition allows to express approximately velocity module and the velocity vector inclination to vertical direction in terms of temperature gradient in explicit form.

\end{abstract}

\section{ Introduction}

A problem of stationary convective flow theoretical description is intriguing but  complicated. The necessity to include both momentum and energy equations with account viscosity and thermoconductivity leads to extra terms in the basic system that do not allow to introduce velocity potential \cite{8}. 
The results of  and experimental study of free convective flows from heating objects are widely published (see \cite{LRDK} and refs therein)  and they are  useful to determine convective heat losses  and tangent forces   by engineers and designers.

The presented paper is devoted to general theoretical study of problem of the description of a stationary two-dimensional flow  near the isothermal surface. 
We consider approximate analytical solution of the equations of a convective flow
induced by an isothermal   body.   The
choice of the coordinate system in the frame of the typical for laminar
natural convection simplifications and for Pr $\thickapprox$ 1 allows to
diminish the number of basic equations.  

As the novel element of the approach we use a transition to coordinates formed by streamlines and spatially built orthogonal lines. It is used further  to simplify the plane version of  Navier-Stokes and Fourier-Kirchoff system. We derive transformation to such coordinates by means of differential geometry, taking into account a necessity to introduce integrating factor that is not equal to unity in a viscous flow. The transition allows to express approximately the mentioned integrating factor, velocity module and the velocity vector  inclination angle with respect  to vertical direction in terms of temperature gradient in explicit form.

The first section contains the basic equations, written similar to \cite{LL,LL1}, the second one defines streamline coordinate system, integrating the equations for orthogonal line via integrating factor introduction. The next section contains a  description of nonsingular perturbation theory that allows to split the system and represent the elements of the novel geometry in terms of the temperature field derivatives. The further section formulate algorithm of the transition of from Cartesian variables to the  streamline coordinate system with an example of velocity field in analytic form. The final section contains an attempt to link the theory with a conventional boundary layer description with all necessary ingredients of the flow sreamlines  geometry. 

\section{The basic equations}

Let us consider a two dimensional stationary flow of incompressible fluid in the gravity field. The flow is generated by a convective heat transfer from
solid plate to the fluid. The plate is isothermal and lies at the half plane
$y\in\lbrack0,\infty).$ We follow the notations of \cite{LL}, writing   the Navier-Stokes
system of equations in the Cartesian coordinates $x,y$%
\begin{equation}
W_{x}\frac{\partial W_{y}}{\partial x}+W_{y}\frac{\partial W_{y}}{\partial
y}=-g\beta\left(  T-T_{\infty}\right)  -\frac{1}{\rho}\frac{\partial
p}{\partial y}+\nu\left(  \frac{\partial^{2}W_{y}}{\partial y^{2}}%
+\frac{\partial^{2}W_{y}}{\partial x^{2}}\right),  \label{NSy}%
\end{equation}%
\begin{equation}
W_{x}\frac{\partial W_{x}}{\partial x}+W_{y}\frac{\partial W_{x}}{\partial
y}=-\frac{1}{\rho}\frac{\partial p}{\partial x}+\nu\left(  \frac{\partial
^{2}W_{x}}{\partial y^{2}}+\frac{\partial^{2}W_{x}}{\partial x^{2}}\right).
\label{NSx}%
\end{equation}

In the above equations the pressure terms are divided in two parts. The first
of them is the hydrostatic one that is equal to mass force $-\rho g$, where
$\rho$ is the density of \ a liquid at the temperature at the non-disturbed
area $T_{\infty}$. The second one $-g\beta\left(  T-T_{\infty}\right)  $ arises
from dependence of the extra density on temperature, $\beta$ is a coefficient
of thermal expansion of the fluid. The last terms of the above equations
represents the friction forces with the kinematic coefficient of viscosity
$\nu.$ I the equations  $W_{x}$ and $W_{y}$ are the components of the fluid velocity
$\overline{W}$ that are shown on the Fig.1; $T$, $p$ - temperature and
pressure disturbances correspondingly.

The mass continuity equation in the conditions of natural convection of
incompressible fluid in the steady state \cite{8} has the form:.%

\begin{equation}
\frac{\partial W_{x}}{\partial x}+\frac{\partial W_{y}}{\partial y}=0.
\label{div}%
\end{equation}

The temperature field is described by the stationary Fourier-Kirchhoff equation:%

\begin{equation}
W_{x}\frac{\partial T}{\partial x}+W_{y}\frac{\partial T}{\partial y}=a\left(
\frac{\partial^{2}T}{\partial y^{2}}+\frac{\partial^{2}T}{\partial x^{2}%
}\right) , \label{FK}%
\end{equation}

After introducing nondimensional variables: $x^{\prime}=x/l,y^{\prime
}=y/l,T^{\prime}=(T-T_{\infty})/\Delta T,p^{\prime}=p/p_{\infty},W_{x}%
^{\prime}=W_{x}/W_{o},$ $W_{y}^{\prime}=W_{y}/W_{o}$ we obtain:%

\begin{equation}
W_{x}^{\prime}\frac{\partial W_{y}^{\prime}}{\partial x^{\prime}}%
+W_{y}^{\prime}\frac{\partial W_{y}^{\prime}}{\partial y^{\prime}}%
=-\frac{g\beta T^{\prime}\Delta Tl}{W_{o}^{2}}-\frac{p_{\infty}}{\rho
W_{o}^{2}}\frac{\partial p^{\prime}}{\partial y^{\prime}}+\nu^{\prime}\left(
\frac{\partial^{2}W_{y}^{\prime}}{\partial y^{\prime2}}+\frac{\partial
^{2}W_{y}^{\prime}}{\partial x^{\prime2}}\right) , \label{NS-1}%
\end{equation}%
\begin{equation}
W_{x}^{\prime}\frac{\partial W_{x}^{\prime}}{\partial x^{\prime}}%
+W_{y}^{\prime}\frac{\partial W_{x}^{\prime}}{\partial y^{\prime}}%
=-\frac{p_{\infty}}{\rho W_{o}^{2}}\frac{\partial p^{\prime}}{\partial
x^{\prime}}+\nu^{\prime}\left(  \frac{\partial^{2}W_{x}}{\partial y^{2}}%
+\frac{\partial^{2}W_{x}}{\partial x^{2}}\right),  \label{NS-2}%
\end{equation}

\begin{equation}
\frac{\partial W_{x}^{\prime}}{\partial x^{\prime}}+\frac{\partial
W_{y}^{\prime}}{\partial y^{\prime}}=0. \label{div'}%
\end{equation}

\begin{equation}
W_{x}^{\prime}\frac{\partial T^{\prime}}{\partial x^{\prime}}+W_{y}^{\prime
}\frac{\partial T^{\prime}}{\partial y^{\prime}}=a^{\prime}\left(
\frac{\partial^{2}T^{\prime}}{\partial y^{\prime2}}+\frac{\partial
^{2}T^{\prime}}{\partial x^{\prime2}}\right),  \label{FK'}%
\end{equation}
where $\frac{\nu}{lW_{o}}=\nu^{\prime},\frac{a}{lW_{o}}=a^{\prime}.$

Next we would formulate the problem of free convection over the heated
inclined isothermal\ plate $x=0,$ $y\in\lbrack0,\infty)$, dropping the primes, see Fig 1.

The form of the continuity equation \eqref{div} allows to introduce the
stream function $\psi$, so as:
\begin{equation}
W_{x}=-\frac{\partial\psi}{\partial y}\text{, \ \ \ \ \ \ \ }W_{y}%
=\frac{\partial\psi}{\partial x}. \label{velo}%
\end{equation}

\section{The streamline coordinate system}

A stream line of the flow is determined by the equation:%
\begin{equation}
\psi(x,y)=n. \label{n}%
\end{equation}
It means that velocity $\overline{W}$ is tangent to the the streamline curve,
inclined to the $x-$ axis by the angle $\theta$.

We introduce tangent $\overline{\tau}$ and normal $\overline{n}$ unit vectors
to the curve (Fig.1), it means that the normal component of the velocity
$W_{n}=0$ and the tangent one $W_{\tau}=W$. We accept in the traditional point
of view thath models real processes on the base of the time independent form
of the streamlines only. Eventual time dependence we would consider as
perturbations with zero mean values.%

\begin{figure}
[ptb]
\begin{center}
\includegraphics[
height=2.3557in,
width=3.3321in
]%
{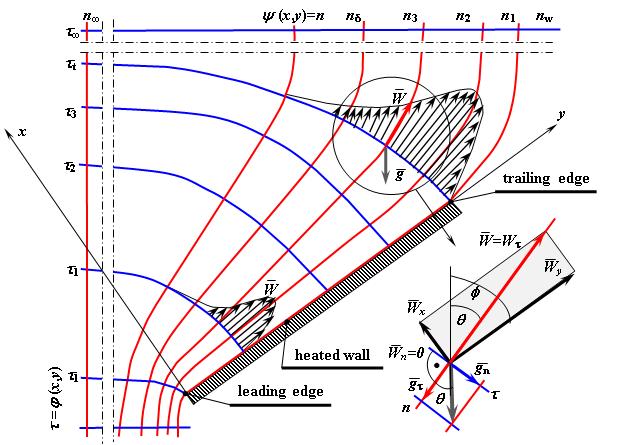}%
\caption{A plate, streamines and orthogonal lines. Velocity components and angle $\theta$.  }
\end{center}
\end{figure}

On the base of the streamlines definition \eqref{n} we have:%
\begin{equation}
y=f(x,n) \label{streamline}%
\end{equation}
and the family of curves to be orthogonal\ to the streamlines:
\begin{equation}
y=h(x,\tau) \label{ortho}%
\end{equation}
we define new curvilinear coordinate system with the variables $(\tau,n)$. The
variables are connected with the Cartesian as:%
\begin{equation}
\tau=\varphi(x,y),\text{ \ \ \ \ }n=\psi(x,y)\text{ \ }. \label{fi}%
\end{equation}
The equation for the function $h$ \eqref{ortho} may be derived from the
equation for a straight line, orthogonal to the streamline \ \eqref{streamline}
in the point $X,Y$ :%
\begin{equation}
\frac{\partial\psi}{\partial y}\left(  X-x\right)  -\frac{\partial\psi
}{\partial x}\left(  Y-y\right)  =0. \label{line}%
\end{equation}

Therefore  the equation for the function $h(x,\tau)$ $\eqref{ortho}$ has the
form (see also  \eqref{velo}  ):%
\begin{equation}
\frac{dh}{dx}=\left[  \frac{\frac{\partial\psi}{\partial y}}{\frac
{\partial\psi}{\partial x}}\right]  _{y=h}=-\cot\theta. \label{h}%
\end{equation}

This differential equation is equivalent to one in the total form%
\begin{equation}
\frac{\partial\psi}{\partial x}dy-\frac{\partial\psi}{\partial y}dx=0
\label{1}%
\end{equation}

The Pfaff form in the l.h.s. of the last equation is exact iff $\frac
{\partial^{2}\psi}{\partial y^{2}}+\frac{\partial^{2}\psi}{\partial x^{2}}=0$,
that means $rot_{z}\overline{W}=0$ such condition strongly restricts the
choice of velocity field (the velocity potential existence which coincides
with $\varphi$) \cite{8}.

It is known that in the two-dimensional case there exist integrating factor
$\mu(x,y)$ to be considered as a new variable of the theory. We identify the
constant of integration of the differential equation  \eqref{h}  with the
variable $\tau$.%
\begin{equation}
\tau=\varphi(x,y)=-\int_{x_{0}}^{x}\mu(x^{\prime},y_{0})\psi_{y}\left(
x^{\prime},y_{0}\right)  dx^{\prime}+\int_{y_{0}}^{y}\mu(x,y^{\prime})\psi
_{x}\left(  x,y^{\prime}\right)  dy^{\prime} \label{tal}%
\end{equation}

The equation that connects the integrating factor $\mu$ and $\psi$ is the
direct corollary of the integrability condition:%
\begin{equation}
\frac{\partial\mu}{\partial x}\frac{\partial\psi}{\partial x}+\frac
{\partial\mu}{\partial y}\frac{\partial\psi}{\partial y}+\mu\Delta\psi=0
\label{mi}%
\end{equation}

The partial derivatives the functions $\psi$ and $\varphi$ $\ref{fi}$
determine the matrix $\widehat{h}$%

\begin{equation}
\widehat{h}= \left(
\begin{array}
[c]{cc}%
\frac{\partial\psi}{\partial x} & \frac{\partial\psi}{\partial y}\\
\frac{\partial\varphi}{\partial x} & \frac{\partial\varphi}{\partial y}%
\end{array}\right)
. \label{lame}%
\end{equation}

The orthogonality condition $\eqref{1}$ yields:%
\begin{equation}
\frac{\partial\psi}{\partial y}\frac{\partial\varphi}{\partial y}%
+\frac{\partial\psi}{\partial x}\frac{\partial\varphi}{\partial x}%
=-\varphi_{y}W_{x}+\varphi_{x}W_{y}=0 \label{ort}%
\end{equation}
and the definition of the stream function $\eqref{velo}$ gives:

$\widehat{h}=%
\left ( \begin{array}
[c]{cc}%
W_{y} & -W_{x}\\
\varphi_{x} & \varphi_{x}\frac{W_{y}}{W_{x}}%
\end{array}\right )
$, $\ \ \ \ \widehat{h}^{-1}=$ $%
\left (\begin{array}
[c]{cc}%
\frac{W_{y}}{W_{x}^{2}+W_{y}^{2}} & \frac{W_{x}^{2}}{W_{x}^{2}\varphi
_{x}+W_{y}^{2}\varphi_{x}}\\
-\frac{W_{x}}{W_{x}^{2}+W_{y}^{2}} & W_{x}\frac{W_{y}}{W_{x}^{2}\varphi
_{x}+W_{y}^{2}\varphi_{x}}%
\end{array}\right )
\allowbreak$

where:%
\begin{equation}
\varphi_{x}=-\mu(x,y_{0})\psi_{y}\left(  x,y_{0}\right)  +\int_{y_{0}}%
^{y}\left(  \mu_{x}(x,y^{\prime})\psi_{x}\left(  x,y^{\prime}\right)
+\mu(x,y^{\prime})\psi_{xx}\left(  x,y^{\prime}\right)  \right)  dy^{\prime}
\label{fix}%
\end{equation}
In the case of $\ \mu=1$ $\ \left(  \Delta\psi=0\right)  $ $,$ $\frac
{\partial\varphi}{\partial x}=\ \varphi_{x}=-\psi_{y}\left(  x,y_{0}\right)
+\int_{y_{0}}^{y}\psi_{xx}\left(  x,y^{\prime}\right)  dy^{\prime}=-\psi
_{y}\left(  x,y_{0}\right)  -\int_{y_{0}}^{y}\psi_{yy}\left(  x,y^{\prime
}\right)  dy^{\prime}=-\psi_{y}\left(  x,y\right)  =W_{x},$ that is equivalent
to the equation introducing velocity potential. This case of $\ rot_{z}%
\overline{W}=0$ means that the terms of viscosity vanish at both Navier-Stokes
equations $\eqref{NSy}$, $\eqref{NSx}$.

In general case of $\mu\neq1$ we propose to consider the integrating factor
($\mu$) as an auxiliary variable.

The components of the metric tensor $G_{ik}=G_{ki}$ of the curvilinear
coordinates are as follows;%

\begin{equation}
G_{nn}=\left(  \frac{\partial x}{\partial n}\right)  ^{2}+\left(
\frac{\partial y}{\partial n}\right)  ^{2}, \label{gnn}%
\end{equation}

\begin{equation}
G_{n\tau}=\frac{\partial x}{\partial n}\frac{\partial x}{\partial\tau}%
+\frac{\partial y}{\partial n}\frac{\partial y}{\partial\tau}, \label{gnt}%
\end{equation}

\begin{equation}
G_{\tau\tau}=\left(  \frac{\partial x}{\partial\tau}\right)  ^{2}+\left(
\frac{\partial y}{\partial\tau}\right)  ^{2}, \label{gtt}%
\end{equation}
with the determinant:%
\begin{equation}
G=G_{nn}G_{\tau\tau}-G_{n\tau}^{2}. \label{g}%
\end{equation}

It is easy to verify that%
\begin{equation}
\widehat{h}^{-1}=%
\left (\begin{array}
[c]{cc}%
\frac{\partial x}{\partial n} & \frac{\partial x}{\partial\tau}\\
\frac{\partial y}{\partial n} & \frac{\partial y}{\partial\tau}%
\end{array}\right )
, \label{g-1}%
\end{equation}
where $x=x(n,\tau)$ and $y=y(n,\tau)$ define the inverse transformation of
$\eqref{fi}.$ Hence the derivatives in $\eqref{g-1}$ and, therefore in $\eqref{lame}%
,$as well as in $\eqref{g}$ are defined by the velocity components (see
$\eqref{velo}$ ). 
Finally the nonzero components of the metric tensor  \eqref{gnn} and
$\eqref{gtt}$:

$G_{nn}=\left(  \frac{W_{y}}{W_{x}^{2}+W_{y}^{2}}\right)  ^{2}+\left(
\frac{W_{x}}{W_{x}^{2}+W_{y}^{2}}\right)  ^{2}=\frac{1}{W^{2}},$

$G_{\tau\tau}=\left(  \frac{W_{x}^{2}}{W_{x}^{2}\varphi_{x}+W_{y}^{2}%
\varphi_{x}}\right)  ^{2}+\left(  W_{x}\frac{W_{y}}{W_{x}^{2}\varphi_{x}%
+W_{y}^{2}\varphi_{x}}\right)  ^{2}=\left(  \frac{W_{x}}{\varphi_{x}}\right)
\allowbreak^{2}\allowbreak\frac{1}{W^{2}},$

with the determinant
\begin{equation}
G=\left(  \frac{W_{x}}{\varphi_{x}}\right)  \allowbreak^{2}\frac{1}{W^{4}},
\label{g1}%
\end{equation}
defines the differential operators of the governing equations in vector form.

The correspondent relations are:

$W_{x}=W\cos\theta,$ \ \ \ $W_{y}=W\sin\theta,$

$\overline{g}=\overline{\tau}g_{\tau}+\overline{n}g_{n}=-\overline{\tau}%
g\sin\theta-\overline{n}g\cos\theta$,

$\nabla p=\frac{\overline{\tau}}{\sqrt{G_{\tau\tau}}}\frac{\partial
p}{\partial\tau}+\frac{\overline{n}}{\sqrt{G_{nn}}}\frac{\partial p}{\partial
n}=\overline{\tau}\frac{\varphi_{x}}{W_{x}}W\frac{\partial p}{\partial\tau
}+\overline{n}W\frac{\partial p}{\partial n}=\overline{\tau}\frac{\varphi_{x}%
}{\cos\theta}\frac{\partial p}{\partial\tau}+\overline{n}W\frac{\partial
p}{\partial n}$,

$\Delta T=\frac{1}{\sqrt{G}}\frac{\partial}{\partial\tau}\left(  \frac
{\sqrt{G}}{G_{\tau\tau}}\frac{\partial T}{\partial\tau}\right)  +\frac
{1}{\sqrt{G}}\frac{\partial}{\partial n}\left(  \frac{\sqrt{G}}{G_{nn}}%
\frac{\partial T}{\partial n}\right)  =\frac{W\varphi_{x}}{\cos\theta}\left[
\frac{\partial}{\partial\tau}\left(  \frac{\varphi_{x}}{W\cos\theta}%
\frac{\partial T}{\partial\tau}\right)  +\frac{\partial}{\partial n}\left(
\frac{W\cos\theta}{\varphi_{x}}\frac{\partial T}{\partial n}\right)  \right]
,$

$\nabla\overline{W}=\frac{1}{\sqrt{G}}\frac{\partial}{\partial\tau}\left(
W_{\tau}\sqrt{\frac{G}{G_{\tau\tau}}}\right)  +\frac{1}{\sqrt{G}}%
\frac{\partial}{\partial n}\left(  W_{n}\sqrt{\frac{G}{G_{nn}}}\right)
=\frac{1}{\sqrt{G}}\frac{\partial}{\partial\tau}\left(  1\right)  =0$,

$\Delta\overline{W}=\nabla\left(  \nabla\overline{W}\right)  -\nabla
\times\left(  \nabla\times\overline{W}\right)  =-\nabla\times\left(
\nabla\times\overline{W}\right)  =\overline{\tau}\frac{\sqrt{G_{\tau\tau}}%
}{\sqrt{G}}\frac{\partial A}{\partial n}-\overline{n}\frac{\sqrt{G_{nn}}%
}{\sqrt{G}}\frac{\partial A}{\partial\tau},$

$W_{x}\frac{\partial T}{\partial x}+W_{y}\frac{\partial T}{\partial y}%
=W_{\tau}\frac{\varphi_{x}}{\cos\theta}\frac{\partial T}{\partial\tau}%
+W_{n}W\frac{\partial T}{\partial n}$,

where: \ $A=\operatorname{rot}_{z}\overline{W}=-\frac{1}{\sqrt{G}}\left(
\frac{\partial\left(  W_{n}\sqrt{G_{nn}}\right)  }{\partial\tau}%
-\frac{\partial\left(  W_{\tau}\sqrt{G_{\tau\tau}}\right)  }{\partial
n}\right)  =\frac{1}{\sqrt{G}}\frac{\partial\left(  W\sqrt{G_{\tau\tau}%
}\right)  }{\partial n}=\frac{W\varphi_{x}}{\cos\theta}\frac{\partial\left(
\frac{W_{x}}{\varphi_{x}}\right)  }{\partial n},$

hence:

$\Delta\overline{W}=\overline{\tau}W\frac{\partial}{\partial n}\left(
\frac{W\varphi_{x}}{\cos\theta}\frac{\partial\left(  \frac{W\cos\theta
}{\varphi_{x}}\right)  }{\partial n}\right)  \allowbreak-\overline{n}%
\frac{\varphi_{x}}{\cos\theta}\frac{\partial}{\partial n}\left(
\frac{W\varphi_{x}}{\cos\theta}\frac{\partial\left(  \frac{W\cos\theta
}{\varphi_{x}}\right)  }{\partial n}\right)  ,$

$\left(  \overline{W}\nabla\right)  \overline{W}=\frac{\nabla W^{2}}%
{2}-\overline{W}\times\left(  \nabla\times\overline{W}\right)  =\frac
{\overline{\tau}}{2}\frac{\varphi_{x}}{\cos\theta}\frac{\partial W^{2}%
}{\partial\tau}+\frac{\overline{n}}{2}W\frac{\partial W^{2}}{\partial
n}-\overline{n}\left(  \frac{\varphi_{x}}{\cos\theta\ \allowbreak}\right)
^{2}W\frac{\partial\left(  \frac{W}{\varphi_{x}}\cos\theta\right)  }{\partial
n},\ $

$-\overline{W}\times\left(  \nabla\times\overline{W}\right)  =\overline
{W}\times\overline{A}=\ -\overline{n}\left(  \frac{\varphi_{x}}{\cos
\theta\ \allowbreak}\right)  ^{2}W\frac{\partial\left(  \frac{W}{\varphi_{x}%
}\cos\theta\right)  }{\partial n}.$

In the new coordinate system the equations $\eqref{NSx},\eqref{NSy}$ and
$\eqref{FK}$ go to the form:%
\begin{equation}
\frac{1}{2}\frac{\partial W^{2}}{\partial\tau}=-\frac{g\beta\Delta
Tl\cos\theta\sin\theta}{W_{o}^{2}\varphi_{x}}T-\frac{p_{\infty}}{\rho
W_{o}^{2}}\frac{\partial p}{\partial\tau}+\nu\frac{W\cos\theta}{\varphi_{x}%
}\frac{\partial}{\partial n}\left(  \frac{W\varphi_{x}}{\cos\theta}%
\frac{\partial\left(  \frac{W\cos\theta}{\varphi_{x}}\right)  }{\partial
n}\right) , \label{NSy1}%
\end{equation}%
\begin{equation}
\frac{1}{2}\frac{\partial W^{2}}{\partial n}-\left(  \frac{\varphi_{x}}%
{\cos\theta}\right)  ^{2}\frac{\partial\left(  \frac{W\cos\theta}{\varphi_{x}%
}\right)  }{\partial n}=-\frac{g\beta\Delta Tl\cos\theta}{WW_{o}^{2}}%
T-\frac{p_{\infty}}{\rho W_{o}^{2}}\frac{\partial p}{\partial n}-\nu
\frac{\varphi_{x}}{W\cos\theta}\frac{\partial}{\partial\tau}\left(
\frac{W\varphi_{x}}{\cos\theta}\frac{\partial\left(  \frac{W\cos\theta
}{\varphi_{x}}\right)  }{\partial n}\right),  \label{NSx1}%
\end{equation}

\begin{equation}
\frac{\partial T}{\partial\tau}=a\left[  \frac{\partial}{\partial\tau}\left(
\frac{\varphi_{x}}{W\cos\theta}\frac{\partial T}{\partial\tau}\right)
+\frac{\partial}{\partial n}\left(  \frac{W\cos\theta}{\varphi_{x}}%
\frac{\partial T}{\partial n}\right)  \right]  \label{FK1}
\end{equation}
\bigskip We have introduced a new variable $\Pi=W^{2}/2+\frac{p}{\rho}%
$\ instead of $p\ $ and cross-differentiate\ $\eqref{NSy1}$, $\eqref{NSx1}$ that yields

$\frac{\partial\left(  \left(  \frac{\varphi_{x}}{\cos\theta\ \allowbreak
}\right)  ^{2}\frac{\partial\left(  \frac{W\cos\theta}{\varphi_{x}}\right)
}{\partial n}-\frac{g\beta\Delta Tl\cos\theta}{WW_{o}^{2}}T-\nu\frac
{\varphi_{x}}{W\cos\theta}\frac{\partial}{\partial\tau}\left(  \frac
{W\varphi_{x}}{\cos\theta}\frac{\partial\left(  \frac{W\cos\theta}{\varphi
_{x}}\right)  }{\partial n}\right)  \right)  }{\partial\tau}=$%

\begin{equation}
\frac{\partial\left(  -\frac{g\beta\Delta Tl\cos\theta\sin\theta}{W_{o}%
^{2}\varphi_{x}}T+\nu\frac{W\cos\theta}{\varphi_{x}}\frac{\partial}{\partial
n}\left(  \frac{W\varphi_{x}}{\cos\theta}\frac{\partial\left(  \frac
{W\cos\theta}{\varphi_{x}}\right)  }{\partial n}\right)  \right)  }{\partial
n}. \label{NScr}%
\end{equation}

So, the problem is formulated on a base of four equations $\eqref{FK1}$,
$\eqref{NScr}$, $\eqref{div}$, $\eqref{mi}$, for three thermodynamical variables
$W$, $\theta$, $T$ and one connected with generalized potential $\varphi$.

Let us underline that the first two equations are already written in new
variables $n$, $\tau$ but the last two in Cartesian ones. Therefore we should
transform them to the same coordinates and the variables: $W$, $\theta.$

The continuity equation $\eqref{div}$ in new variables yields

$\frac{\partial W\cos\theta}{\partial x}+\frac{\partial W\sin\theta}{\partial
y}=\frac{\partial W}{\partial x}\cos\theta-\frac{\partial\theta}{\partial
x}W\sin\theta+\frac{\partial W}{\partial y}\sin\theta+\frac{\partial\theta
}{\partial y}W\cos\theta=\left(  \overset{-}{\tau},\nabla W\right)  +W\left(
\overset{-}{n},\nabla\theta\right)=0,$
where:%
\begin{equation}
\overset{-}{\tau}=\left(  \cos\theta,\sin\theta\right)  ,\overset{-}%
{n}=\left(  -\sin\theta,\cos\theta\right)  . \label{taun}%
\end{equation}
and in curvilinear coordinates the gradients are: \ %

\begin{equation}
\nabla W=\frac{\overline{\tau}}{\sqrt{G_{\tau\tau}}}\frac{\partial W}%
{\partial\tau}+\frac{\overline{n}}{\sqrt{G_{nn}}}\frac{\partial W}{\partial
n}=\overline{\tau}\frac{\varphi_{x}}{W_{x}}W\frac{\partial W}{\partial\tau
}+\overline{n}W\frac{\partial W}{\partial n}=\overline{\tau}\frac{\varphi_{x}%
}{\cos\theta}\frac{\partial W}{\partial\tau}+\overline{n}W\frac{\partial
W}{\partial n}, \label{gradw}%
\end{equation}
and \ %
\begin{equation}
\nabla\theta=\overline{\tau}\frac{\varphi_{x}}{\cos\theta}\frac{\partial
\theta}{\partial\tau}+\overline{n}W\frac{\partial\theta}{\partial n}.
\label{gradteta}%
\end{equation}
That finally gives
\begin{equation}
\frac{\varphi_{x}}{\cos\theta}\frac{\partial W}{\partial\tau}+W^{2}%
\frac{\partial\theta}{\partial n}=0 .\label{div-1}%
\end{equation}
The last equation of integrability $\eqref{mi}$ reads

$\frac{\partial\mu}{\partial x}\frac{\partial\psi}{\partial x}+\frac
{\partial\mu}{\partial y}\frac{\partial\psi}{\partial y}+\mu\Delta\psi
=\frac{\partial\mu}{\partial x}W\sin\theta-\frac{\partial\mu}{\partial y}%
W\cos\theta+\mu\left(  \frac{\partial^{2}\psi}{\partial y^{2}}+\frac
{\partial^{2}\psi}{\partial x^{2}}\right)  =\frac{\partial\mu}{\partial
x}W\sin\theta-\frac{\partial\mu}{\partial y}W\cos\theta+$

$+\mu\left(  -\frac{\partial W\cos\theta}{\partial y}+\frac{\partial
W\sin\theta}{\partial x}\right)  =\frac{\partial\left(  \mu W\sin
\theta\right)  }{\partial x}-\frac{\partial\left(  \mu W\cos\theta\right)
}{\partial y}=\operatorname{rot}_{z}\left(  \overline{W}\mu\right)  =0.$

In new variables the $z$- component of the operator curl takes the form%
\begin{equation}
\operatorname{rot}_{z}\left(  \overline{W}\mu\right)  =-\frac{1}{\sqrt{G}%
}\left(  \frac{\partial\left(  \mu W_{n}\sqrt{G_{nn}}\right)  }{\partial\tau
}-\frac{\partial\left(  \mu W_{\tau}\sqrt{G_{\tau\tau}}\right)  }{\partial
n}\right)  =\frac{\partial\left(  \frac{\mu W\cos\theta}{\varphi_{_{x}}%
}\right)  }{\partial n}=0. \label{rotmi}%
\end{equation}
The last relation for assumption $\frac{\mu W\cos\theta}{\varphi_{_{x}}}=1$
gives the link between $\mu$ and $\varphi_{_{x}}$:%
\begin{equation}
\mu=\frac{\varphi_{_{x}}}{W\cos\theta}. \label{mifi}%
\end{equation}
The constant of integration of $\eqref{rotmi}$ is chosen as unit on the base
of the freedom in integration factor.

Returning to the equation $\ref{rotmi}$ in vector form and using
($\ref{taun}$) one can rewrite it as

$-\frac{1}{\mu}\frac{\partial\mu}{\partial x}\sin\theta+\frac{1}{\mu}%
\frac{\partial\mu}{\partial y}\cos\theta-\frac{1}{W}\left(  -\frac{\partial
W\cos\theta}{\partial y}+\frac{\partial W\sin\theta}{\partial x}\right)  =0$%

\begin{equation}
-\left(  \overline{n},\nabla\ln\mu\right)  -\left(  \overline{n},\nabla\ln
W\right)  +\left(  \overline{\tau},\nabla\theta\right)  =0, \label{IV}%
\end{equation}
on the base of  $\eqref{gradteta}$, having in mind $\nabla\ln W=\frac{\nabla
W}{W}$. Finally, in the curvilinear coordinates
\begin{equation}
-\frac{\partial\left(  \ln\mu W\right)  }{\partial n}+\mu\frac{\partial\theta
}{\partial\tau}=0. \label{IVb}%
\end{equation}

Let us plug the relation  $\eqref{mifi}$  into $\eqref{div-1}$ arriving at%
\begin{equation}
\mu\frac{\partial\left(  \ln W\right)  }{\partial\tau}+\frac{\partial\theta
}{\partial n}=0, \label{V}%
\end{equation}

and, next%
\begin{equation}
\mu=-\frac{\frac{\partial\theta}{\partial n}}{\frac{\partial\left(  \ln
W\right)  }{\partial\tau}}. \label{mi-V}%
\end{equation}

To solve the problem for the equations $\eqref{NScr},\eqref{FK1}$ and $\eqref{IVb}$
in which $\eqref{mi-V}$ is implied.

Transforming the N-S equations  one has

$\frac{\partial\left(  \left(  \mu W\right)  ^{2}\frac{\partial\left(
\frac{1}{\mu}\right)  }{\partial n}-\frac{g\beta\Delta Tl\cos\theta}%
{WW_{o}^{2}}T-\nu\mu\frac{\partial}{\partial\tau}\left(  W^{2}\mu
\frac{\partial\left(  \frac{1}{\mu}\right)  }{\partial n}\right)  \right)
}{\partial\tau}=\frac{\partial\left(  -W^{2}\frac{\partial\mu}{\partial
n}-\frac{g\beta\Delta Tl\cos\theta}{WW_{o}^{2}}T+\nu\mu\frac{\partial
}{\partial\tau}\left(  W^{2}\frac{1}{\mu}\frac{\partial\mu}{\partial
n}\right)  \right)  }{\partial\tau},$

$\frac{\partial\left(  -\frac{g\beta\Delta Tl\sin\theta}{\mu WW_{o}^{2}}%
T+\nu\frac{1}{\mu}\frac{\partial}{\partial n}\left(  W^{2}\mu\frac
{\partial\left(  \frac{1}{\mu}\right)  }{\partial n}\right)  \right)
}{\partial n}=\frac{\partial\left(  -\frac{g\beta\Delta Tl\sin\theta}{\mu
WW_{o}^{2}}T-\nu\frac{1}{\mu}\frac{\partial}{\partial n}\left(  W^{2}\frac
{1}{\mu}\frac{\partial\mu}{\partial n}\right)  \right)  }{\partial n},$

 one arrives at%
\begin{equation}
\frac{\partial\left(  -W^{2}\frac{\partial\mu}{\partial n}-\frac{g\beta\Delta
Tl\cos\theta}{WW_{o}^{2}}T+\nu\mu\frac{\partial}{\partial\tau}\left(
W^{2}\frac{\partial\left(  \ln\mu\right)  }{\partial n}\right)  \right)
}{\partial\tau}=\frac{\partial\left(  -\frac{g\beta\Delta Tl\sin\theta}{\mu
WW_{o}^{2}}T-\nu\frac{1}{\mu}\frac{\partial}{\partial n}\left(  W^{2}%
\frac{\partial\left(  \ln\mu\right)  }{\partial n}\right)  \right)  }{\partial
n}. \label{N-S-2}%
\end{equation}

 Introducing the Rayleigh number $Ra=\frac{g\beta\Delta Tl^{3}}{\nu a}$ we obtained%
\begin{equation}
\frac{\partial\left(  -\frac{W^{2}}{\nu}\frac{\partial\mu}{\partial
n}-aRa\frac{T\cos\theta}{W}+\mu\frac{\partial}{\partial\tau}\left(  W^{2}%
\frac{\partial\left(  \ln\mu\right)  }{\partial n}\right)  \right)  }%
{\partial\tau}=\frac{\partial\left(  -aRa\frac{T\sin\theta}{\mu W}-\frac
{1}{\mu}\frac{\partial}{\partial n}\left(  W^{2}\frac{\partial\left(  \ln
\mu\right)  }{\partial n}\right)  \right)  }{\partial n} \label{N-S-3}%
\end{equation}

The equation $\eqref{FK1}$ is transformed as:
\begin{equation}
\frac{\partial T}{\partial\tau}=a\left[  \frac{\partial}{\partial\tau}\left(
\mu\frac{\partial T}{\partial\tau}\right)  +\frac{\partial}{\partial n}\left(
\frac{1}{\mu}\frac{\partial T}{\partial n}\right)  \right] . \label{FK3}%
\end{equation}

The equations $\eqref{N-S-2}$, $\eqref{FK1}$ and%
\begin{equation}
-\frac{\partial\left(  \ln\mu W\right)  }{\partial n}+\mu\frac{\partial\theta
}{\partial\tau}=0. \label{div-3}%
\end{equation}
together with the expression for $\mu$  $\eqref{mi-V}$  form the system of
three equations for three variables $W$, $\theta$ and $T$ that is equivalent to
the basic one. This system, with boundary conditions acount, we consider as
the formulation of the problem to be solved in new independent coordinates
$n,$ $\tau.$

We should formulate boundary conditions . It is helpful to draw the coordinate
system in terms of new variables and sketch the unknown function behavior as, for example at  Fig.2
 
\begin{figure}
[ptb]
\begin{center}
\includegraphics[
height=2.3557in,
width=3.3321in
]%
{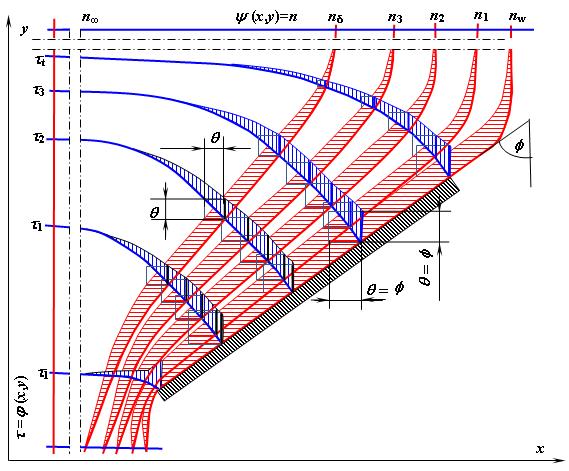}%
\label{Fig2}
\caption{The sketch of $\theta$ profiles along streamlines an along orthogonal ones. }
\end{center}
\end{figure}
 
\section{Nonsingular perturbation theory}

We apply the nonsingular perturbation  theory to the system of equations
$\eqref{N-S-2}$, $\eqref{FK1}$ and $\eqref{IVb}$ in which a small parameters indicate
slow changes of dynamic variables as a function of the correspondent
independent variable \cite{13}. Let the variables depend on small parameter
$\varepsilon$ \cite{4}.

Transport of mass of fluid particles take place along the stream lines $\tau$
($n=const$). It means that the main contribution to the heat transport is
realized by such particles. Then the gradient component of temperature
$\partial T/\partial n$ is large compared to $\partial T/\partial\tau.$ Moreover
the heat exchange between fluid particles at neighbor stream lines is defined
by thermal conductivity that is characterized by second derivative by $\tau$.
This assumption may be expressed by small parameter $\varepsilon$ introduction
in temperature field as:%
\begin{equation}
T=T\left(  n,\varepsilon^{2}\tau\right)  . \label{T}%
\end{equation}
The transport of the fluid particles momentum is similar but it is determined
by buoyancy and viscosity forces that act in different directions. The module
of velocity is changed essentially along the perpendicular direction to stream
lines, while its angle of inclination changes opposite:%
\begin{equation}
W=W\left(  n,\varepsilon\tau\right)  ,\ \ \ \ \ \theta=\theta\left(
\varepsilon n,\tau\right) . \ \ \label{epsi}%
\end{equation}
 \begin{figure}
[ptb]
\begin{center}
\includegraphics[
height=2.3557in,
width=3.3321in
]%
{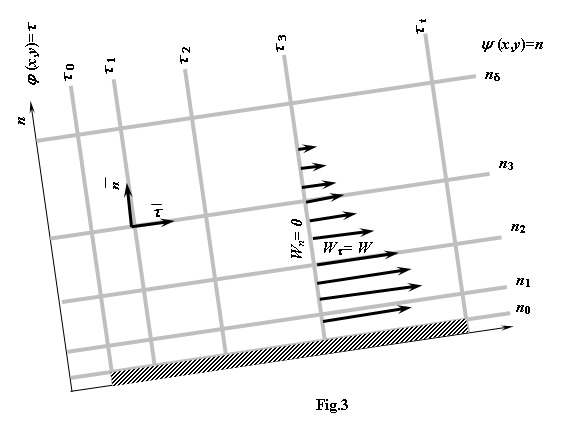} 
\label{Fig.3}
\caption{The sketch of $W$ profiles across streamlines in new coordinates.}
\end{center}
\end{figure}

The heat transfer equation $\eqref{FK1}$ in the first order of the parameter
$\varepsilon^{2}$ gives%
\begin{equation}
\frac{\partial T\left(  n,\tau\right)  }{\partial\tau}=a\left[  \frac
{\partial}{\partial n}\left(  \frac{1}{\mu}\frac{\partial T\left(
n,\tau\right)  }{\partial n}\right)  \right] . \label{FK2}%
\end{equation}

After approximations the parameter is chosen conventionally as $\varepsilon
=1.$ by $n$ it is solved with respect to $\mu$%
\begin{equation}
\mu=\frac{a\frac{\partial T\left(  n,\tau\right)  }{\partial n}}{\int
\frac{\partial T\left(  n,\tau\right)  }{\partial\tau}dn}. \label{mi6}%
\end{equation}

$\frac{\partial T\left(  n,\tau\varepsilon^{2}\right)  }{\partial\tau
}=\allowbreak\varepsilon^{2}D_{2}T\left(  n,\tau\varepsilon^{2}\right)
=\allowbreak\varepsilon^{2}D_{2}T\left(  n,\tau\varepsilon^{2}\right),  $

$\frac{\partial^{2}T\left(  n,\tau\varepsilon^{2}\right)  }{\partial\tau^{2}%
}=\allowbreak\varepsilon^{4}D_{2,2}T\left(  n,\tau\varepsilon^{2}\right), $

$\frac{\partial^{2}W\left(  n,\varepsilon\tau\right)  }{\partial n\partial
\tau}=\allowbreak\varepsilon D_{1,2}W\left(  n,\tau\varepsilon\right). $

The continuity equation $\eqref{div-3}$ with $\eqref{mi6}$ yields :%
\begin{equation}\begin{array}{c}
\frac{\partial\theta\left(  n,\tau\right)  }{\partial n}^{2}\frac
{\partial\theta\left(  n,\tau\right)  }{\partial\tau}\left(  W\left(
n,\tau\right)  \right)  ^{2}-\frac{\partial\theta\left(  n,\tau\right)
}{\partial n}W\left(  n,\tau\right)  \frac{\partial^{2}W\left(  n,\tau\right)
}{\partial n\partial\tau}+\\\frac{\partial W\left(  n,\tau\right)  }%
{\partial\tau}W\left(  n,\tau\right)  \frac{\partial^{2}\theta\left(
n,\tau\right)  }{\partial n\partial n}+2\frac{\partial\theta\left(
n,\tau\right)  }{\partial n}\frac{\partial W\left(  n,\tau\right)  }{\partial
n}\frac{\partial W\left(  n,\tau\right)  }{\partial\tau}=0.
\end{array}
\end{equation}
In the first approximation we write%
\begin{equation}
-W\left(  n,\tau\right)  \frac{\partial^{2}W\left(  n,\tau\right)  }{\partial
n\partial\tau}+2\frac{\partial W\left(  n,\tau\right)  }{\partial n}%
\frac{\partial W\left(  n,\tau\right)  }{\partial\tau}+\frac{\partial
\theta\left(  n,\tau\right)  }{\partial n}\frac{\partial\theta\left(
n,\tau\right)  }{\partial\tau}\left(  W\left(  n,\tau\right)  \right)  ^{2}=0.
\label{cont}%
\end{equation}

From the equalities $\eqref{mi6}$ and $\eqref{div-3}$ we have

$-\frac{\partial\left(  \ln\left(  \frac{a\frac{\partial T\left(
n,\tau\right)  }{\partial n}}{\int\frac{\partial T\left(  n,\tau\right)
}{\partial\tau}dn}\right)  W\right)  }{\partial n}+\frac{a\frac{\partial
T\left(  n,\tau\right)  }{\partial n}}{\int\frac{\partial T\left(
n,\tau\right)  }{\partial\tau}dn}\frac{\partial\theta\left(  n,\tau\right)
}{\partial\tau}=0,$

$\frac{\partial\left(  \ln\left(  a\frac{\partial T\left(  n,\tau\right)
}{\partial n}\right)  +\ln\left(  W\left(  n,\tau\right)  \right)  -\ln\left(
\int\frac{\partial T\left(  n,\tau\right)  }{\partial\tau}dn\right)  \right)
}{\partial n}=$

$\frac{1}{\frac{\partial T\left(  n,\tau\right)  }{\partial n}}\frac
{\partial^{2}T\left(  n,\tau\right)  }{\partial n\partial n}-\frac
{\frac{\partial T\left(  n,\tau\right)  }{\partial\tau}}{\int\frac{\partial
T\left(  n,\tau\right)  }{\partial\tau}\,dn}+\frac{\frac{\partial W\left(
n,\tau\right)  }{\partial n}}{W\left(  n,\tau\right)  },\allowbreak$%

\begin{equation}
\frac{1}{\frac{\partial T\left(  n,\tau\right)  }{\partial n}}\frac
{\partial^{2}T\left(  n,\tau\right)  }{\partial n\partial n}-\frac
{\frac{\partial T\left(  n,\tau\right)  }{\partial\tau}}{\int\frac{\partial
T\left(  n,\tau\right)  }{\partial\tau}\,dn}+\frac{\frac{\partial W\left(
n,\tau\right)  }{\partial n}}{W\left(  n,\tau\right)  }+\frac{a\frac{\partial
T\left(  n,\tau\right)  }{\partial n}}{\int\frac{\partial T\left(
n,\tau\right)  }{\partial\tau}dn}\frac{\partial\theta\left(  n,\tau\right)
}{\partial\tau}=0. \label{div-IV}%
\end{equation}

Equalizing the expressions $\eqref{mi-V}$ and $\eqref{mi6}$ for $\mu$
yields$\ \ $%
\begin{equation}
-\frac{\frac{\partial\theta\left(  n,\tau\right)  }{\partial n}}%
{\frac{\partial\left(  \ln W\left(  n,\tau\right)  \right)  }{\partial\tau}%
}=\frac{a\frac{\partial T\left(  n,\tau\right)  }{\partial n}}{\int
\frac{\partial T\left(  n,\tau\right)  }{\partial\tau}dn}, \label{mi=mi}%
\end{equation}
or using shorthands for derivatives: $\ \ \ \ \ $
$\ $%
\begin{equation}
\ln_{\tau}W=-\frac{\theta_{n}}{aT_{n}}\int T_{_{\tau}}\,dn. \label{mi-mi1}%
\end{equation}
Integrating by $\tau$ we have%
\begin{equation}
W=e^{-\int\frac{\partial\theta}{\partial n}\frac{\int\frac{\partial T\left(
n,\varepsilon\tau\right)  }{\partial\tau}dn}{a\frac{\partial T\left(
n,\varepsilon\tau\right)  }{\partial n}}d\tau}. \label{mi-mi2}%
\end{equation}

Let us analyse the contribution of the third term of $\eqref{div-IV}$ on the base of $\ref{mi-mi2}$

$\frac{\frac{\partial W\left(  n,\tau\right)  }{\partial n}}{W\left(
n,\tau\right)  }=\frac{\partial\left(  \ln W\right)  }{\partial n}%
=\frac{\partial\left(  -\int\frac{\partial\theta\left(  n,\tau\right)
}{\partial n}\frac{\int\frac{\partial T\left(  n,\tau\right)  }{\partial\tau
}dn}{a\frac{\partial T\left(  n,\tau\right)  }{\partial n}}d\tau\right)
}{\partial n}=\allowbreak$

$-\frac{1}{a}\int\left(  \frac{1}{\frac{\partial T\left(  n,\tau\right)
}{\partial n}}\frac{\partial T\left(  n,\tau\right)  }{\partial\tau}%
\frac{\partial\theta\left(  n,\tau\right)  }{\partial n}+\frac{1}%
{\frac{\partial T\left(  n,\tau\right)  }{\partial n}}\int\frac{\partial
T\left(  n,\tau\right)  }{\partial\tau}\,dn\frac{\partial^{2}\theta\left(
n,\tau\right)  }{\partial n\partial n}-\frac{1}{\frac{\partial T\left(
n,\tau\right)  }{\partial n}^{2}}\frac{\partial\theta\left(  n,\tau\right)
}{\partial n}\int\frac{\partial T\left(  n,\tau\right)  }{\partial\tau
}\,dn\frac{\partial^{2}T\left(  n,\tau\right)  }{\partial n\partial n}\right).
\,d\tau\allowbreak\allowbreak$

From basic estimations expressed by small parameters entrance at $\eqref{T}$ and
$\eqref{epsi}$ it follows that all terms are of the order $\varepsilon^{^{2}}.$
Hence the equation ($ \eqref{div-IV}$) is reduced to:%
\begin{equation}
\frac{1}{\frac{\partial T\left(  n,\tau\right)  }{\partial n}}\frac
{\partial^{2}T\left(  n,\tau\right)  }{\partial n\partial n}+\frac
{\frac{\partial T\left(  n,\tau\right)  }{\partial n}}{\int\frac{\partial
T\left(  n,\tau\right)  }{\partial\tau}dn}\left(  a\frac{\partial\theta\left(
n,\tau\right)  }{\partial\tau}-1\right)  =0. \label{Tteta}%
\end{equation}

The result allows to express the field of velocity angles $\theta\left(
n,\tau\right)  $ as a function of temperature field $T\left(  n,\tau\right)
.$%
\begin{equation}
\theta\left(  n,\tau\right)  =\int\left(  -\frac{\int\frac{\partial T\left(
n,\tau\right)  }{\partial\tau}dn}{a\left(  \frac{\partial T\left(
n,\tau\right)  }{\partial n}\right)  ^{2}}\frac{\partial^{2}T\left(
n,\tau\right)  }{\partial n\partial n}+\frac{1}{a}\right)  d\tau.\label{Teta}%
\end{equation}

Under such assumptions the Navier-Stokes equations $ \eqref{N-S-2}$ in first
approximation with respect to the small parameter $\varepsilon$ may derived
in similar way. We however in this work concentrate our efforts on the problem
of velocity field determination on the base of temperature fields by means of
equations $ \eqref{mi-mi2}$ and  $ \eqref{Tteta}$, see the typical $W,\theta,T$ profiles across and along streamlines in new coordinates at Fig.4.
 \begin{figure}
[ptb]
\begin{center}
\includegraphics[
height=2.3557in,
width=3.3321in
]%
{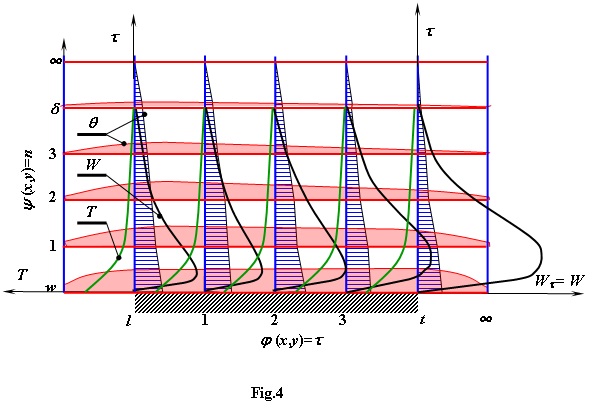} 
\label{Fig.4}
\caption{The sketch of $W,\theta,T$ profiles across and along streamlines in new coordinates.}
\end{center}
\end{figure}

\section{On the algorithm of transition from Cartesian to coordinates $n,\tau$
and vice versa}

Let us recall that $W_{x}=-\frac{\partial\psi}{\partial y}$ and $W_{y}%
=\frac{\partial\psi}{\partial x}$, then

$dn=\frac{\partial\psi}{\partial y}dx+\frac{\partial\psi}{\partial x}%
dy=W_{y}dx-W_{x}dy,$

where $W_{x}=W\cos\theta$ and $W_{y}=W\sin\theta,$  while by the definition of
velocity potential

$d\tau=\frac{\partial\varphi}{\partial x}dx+\frac{\partial\varphi}{\partial
y}dy.$

From  \eqref{ort}  it follows $-\varphi_{y}W_{x}+\varphi_{x}W_{y}=0.$

Next, ( $ \ref{mifi}$) and ( $ \ref{mi6}$) give $\mu=\frac{\varphi_{_{x}}}%
{W\cos\theta}=\frac{a\frac{\partial T\left(  n,\tau\right)  }{\partial n}%
}{\int\frac{\partial T\left(  n,\tau\right)  }{\partial\tau}dn},$ that allow
to express $\varphi_{_{x}}$%
\begin{equation}
\frac{\partial\varphi}{\partial x}=\varphi_{_{x}}=W\cos\theta\frac
{a\frac{\partial T\left(  n,\tau\right)  }{\partial n}}{\int\frac{\partial
T\left(  n,\tau\right)  }{\partial\tau}dn}, \label{fix1}%
\end{equation}
and%
\begin{equation}
\frac{\partial\varphi}{\partial y}=\varphi_{y}=\frac{\varphi_{x}W_{y}}{W_{x}%
}=\varphi_{x}\tan\theta, \label{fiy1}%
\end{equation}
where $W$ is evaluated from ( $ \ref{mi-mi2}$) and $\theta$ is found from ($ \ref{Teta}$)%
\begin{equation}
dn=Adx+Bdy,\ \ \ \ \ d\tau=Cdx+Gdy. \label{dndtau}%
\end{equation}

The coefficients are%
\begin{equation}%
\begin{tabular}
[c]{l}%
$A=W\sin(\theta),$\\
$B=$ $-W\cos\left(  \theta\right),  $\\
$C=W\cos\left(  \theta\right)  \frac{a\frac{\partial T\left(  n,\tau\right)
}{\partial n}}{\int\frac{\partial T\left(  n,\tau\right)  }{\partial\tau}%
dn}=\mu W\cos\left(  \theta\right),  $\\
$G=W\sin\left(  \theta\right)  \frac{a\frac{\partial T\left(  n,\tau\right)
}{\partial n}}{\int\frac{\partial T\left(  n,\tau\right)  }{\partial\tau},%
dn}=\mu W\sin\left(  \theta\right).  $%
\end{tabular}
\label{ABCG}%
\end{equation}
For differentials in Cartesian coordinates we arrive at
\begin{equation}%
\begin{tabular}
[c]{l}%
$dy=-\frac{Ad\tau-Cdn}{BC-AG}=\frac{1}{\mu W}\left(  \sin\left(
\theta\right)  d\tau-\mu\cos\left(  \theta\right)  dn\right),  $\\
$dx=\frac{Bd\tau-Gdn}{BC-AG}=\frac{1}{\mu W}\left(  \cos\left(  \theta\right)
d\tau+\mu\sin(\theta)dn\right).  $%
\end{tabular}
\label{dxdy}%
\end{equation}

$dy=-\frac{Ad\tau-Cdn}{BC-AG}=\frac{1}{\mu
W}\left(  \sin\left(  \theta\right)  d\tau-\mu\cos\left(  \theta\right)
dn\right)  \allowbreak$

$dx=\frac{Bd\tau-Gdn}{BC-AG}=\frac{1}{\mu
W}\left(  \cos\left(  \theta\right)  d\tau+\mu\sin(\theta)dn\right)  $

In conditions $d\tau=0$:

$dy=-\frac{Ad\tau-Cdn}{BC-AG}=\frac{1}{W}\left(  -\cos\left(  \theta\right)
dn\right)  $ $\ dn=-\frac{Wdy}{\cos\left(  \theta\right),  }$

$dx=\frac{1}{W}\left(  \sin(\theta)dn\right),  $ \ \ $dn=\frac{Wdy}{\sin\left(
\theta\right)  }.$

The gradients of temperature and velocity module are rescaled as 

$\frac{dT}{dn}=\frac{\sin(\theta)}{W}\frac{dT}{dx},$ $\ \ \ \ \ \ \frac{dT}%
{dn}=-\frac{\cos(\theta)}{W}\frac{dT}{dy},$

$\frac{dW}{dn}=\frac{\sin(\theta)}{W}\frac{dW}{dx},$ $\ \ \ \ \ \ \frac{dW}%
{dn}=-\frac{\cos(\theta)}{W}\frac{dW}{dy}.$

\bigskip If the function $W$ exponentially decays at a vicinity of the layer
boundary, asymptotically we have:

$W=W_{0}e^{-\alpha x},\frac{dW}{dx}=-\alpha W_{0}e^{-\alpha x,}$

\bigskip the gradient by $n$-variable do not decay:

$\frac{\sin(\theta)}{W}\frac{dW}{dx}=-\frac{\sin(\theta)}{W_{0}e^{-\alpha x}%
}\alpha W_{0}e^{-\alpha x}=-\alpha\sin(\theta.)$

Similar behavior demonstrates the temperature
 \begin{figure}
[ptb]
\begin{center}
\includegraphics[
height=2.3557in,
width=3.3321in
]%
{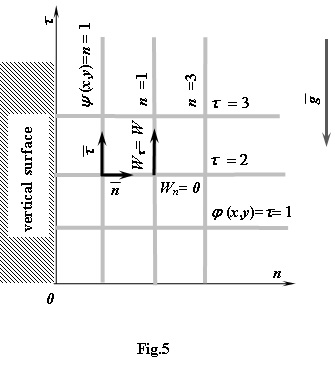} 
\label{Fig.5}
\caption{The streamlines coordinates for the vertical plate: on a way to discretization.}
\end{center}
\end{figure}

Next, if $dn=0,$

$dy=\ \frac{1}{\mu W}\sin\left(  \theta\right)  d\tau,$

$dx=\ \frac{1}{\mu W}\cos\left(  \theta\right)  d\tau.$

We have similar relations, with account of $\mu$ behavior

\bigskip$\frac{dT}{d\tau}=\frac{dT}{dx}\frac{\mu W}{\cos\left(  \theta\right)
}=\frac{dT}{dx}\frac{\mu W}{\sin\left(  \theta\right)  }.$

This is the nonlinear system with respect to that gives the link between the temperature  gradients components in both coordinate systems. A direct application is possible in its discrete version (see Fig.5 ).

\section{The boundary layer approximation}

In the theory of boundary layer the models for temperature and velocity fields
are expressed as \cite{8,12}: $\left(  T-T_{\infty}\right)  =\Delta T(1-x/ \delta
(y))^{2},$ where $ \delta (y)$ is the boundary layer thickness.

By the similarity of descriptions in both coordinate system we assume:

$\left(  T-T_{\infty}\right)  =\left(  T_{w}-T_{\infty}\right)  (1-n/ \delta
(\tau))^{2}.$

Such model is based on the concept of boundary layer. It means that boundary
layer thickness $ \delta (\tau)$ is defined by the equation of layer boundary
$n= \delta (\tau)$ with condition $W$ $\sim0$ at the boundary and the angle
$\theta( \delta (\tau),\tau)$\ determines the form of the boundary curve (see
Fig.4).

Looking for the necessary elements of geometrical description we should write
the coefficients from  \eqref{ABCG} we start from   \eqref{Teta}:

$\theta\left(  n,\tau\right)  =\int\left(  -\frac{\int\frac{\partial T\left(
n,\tau\right)  }{\partial\tau}dn}{a\left(  \frac{\partial T\left(
n,\tau\right)  }{\partial n}\right)  ^{2}}\frac{\partial^{2}T\left(
n,\tau\right)  }{\partial n\partial n}+\frac{1}{a}\right)  d\tau.$

Evaluating $\int\frac{\partial T\left(  n,\tau\right)  }{\partial\tau}dn,$ one have

$\left(  T_{w}-T_{\infty}\right)  \frac{\partial\left(  (1-n/ \delta
(\tau))^{2}\right)  }{\partial\tau}=\allowbreak-2\frac{n}{\left(   \delta
\left(  \tau\right)  \right)  ^{3}}\left(  T_{w}-T_{\infty}\right)
\frac{\partial \delta \left(  \tau\right)  }{\partial\tau}\left(  n- \delta
\left(  \tau\right)  \right),  \allowbreak$

$\allowbreak\int\frac{\partial T\left(  n,\tau\right)  }{\partial\tau}%
dn=\int\left(  -2\frac{n}{\left(   \delta \left(  \tau\right)  \right)  ^{3}%
}\left(  T_{w}-T_{\infty}\right)  \frac{\partial \delta \left(  \tau\right)
}{\partial\tau}\left(  n- \delta \left(  \tau\right)  \right)  \right)
dn=\allowbreak\frac{1}{3}\frac{n^{2}}{\left(   \delta \left(  \tau\right)
\right)  ^{3}}\left(  T_{w}-T_{\infty}\right)  \left(  3 \delta \left(
\tau\right)  -2n\right)  \frac{\partial \delta \left(  \tau\right)  }%
{\partial\tau},\allowbreak$

$\left(  \frac{\partial T\left(  n,\tau\right)  }{\partial n}\right)
=\frac{\partial\left(  \left(  T_{w}-T_{\infty}\right)  (1-n/ \delta
(\tau))^{2}\right)  }{\partial n}=\allowbreak-\frac{1}{\left(   \delta \left(
\tau\right)  \right)  ^{2}}\left(  T_{w}-T_{\infty}\right)  \left(  2 \delta
\left(  \tau\right)  -2n\right),  $

$\frac{\partial^{2}\left(  \left(  T_{w}-T_{\infty}\right)  (1-n/ \delta
(\tau))^{2}\right)  }{\partial n\partial n}=\allowbreak\frac{2}{\left(
 \delta \left(  \tau\right)  \right)  ^{2}}\left(  T_{w}-T_{\infty}\right),  $

$-\frac{\int\frac{\partial T\left(  n,\tau\right)  }{\partial\tau}dn}{a\left(
\frac{\partial T\left(  n,\tau\right)  }{\partial n}\right)  ^{2}}%
\frac{\partial^{2}T\left(  n,\tau\right)  }{\partial n\partial n}+\frac{1}%
{a}=-\frac{\left(  \frac{1}{3}\frac{n^{2}}{\left(   \delta \left(
\tau\right)  \right)  ^{3}}\left(  T_{w}-T_{\infty}\right)  \left(  3 \delta
\left(  \tau\right)  -2n\right)  \frac{\partial \delta \left(  \tau\right)
}{\partial\tau}\right)  }{a\left(  -\frac{1}{\left(   \delta \left(
\tau\right)  \right)  ^{2}}\left(  T_{w}-T_{\infty}\right)  \left(  2 \delta
\left(  \tau\right)  -2n\right)  \right)  ^{2}}\left(  \frac{2}{\left(
 \delta \left(  \tau\right)  \right)  ^{2}}\left(  T_{w}-T_{\infty}\right)
\right)  +\frac{1}{a}=\frac{1}{a}-\frac{2}{3a}\frac{n^{2}}{ \delta \left(
\tau\right)  \left(  2 \delta \left(  \tau\right)  -2n\right)  ^{2}}\left(
3 \delta \left(  \tau\right)  -2n\right)  \allowbreak\frac{\partial \delta
\left(  \tau\right)  }{\partial\tau},$

having

$\allowbreak\theta\left(  n,\tau\right)  =\allowbreak\frac{1}{a}-\frac{n^{2}%
}{6a}\frac{1}{ \delta \left(  \tau\right)  \left(   \delta \left(
\tau\right)  -n\right)  ^{2}}\allowbreak\left(  3 \delta \left(  \tau\right)
-2n\right)  \frac{\partial \delta \left(  \tau\right)  }{\partial\tau.}$

Differentiating yields: 

$\frac{\partial\theta}{\partial n}=\frac{\partial\left(  \frac{1}{a}-\frac
{2}{3a}\frac{n^{2}}{ \delta \left(  \tau\right)  \left(  2 \delta \left(
\tau\right)  -2n\right)  ^{2}}\allowbreak\left(  3 \delta \left(  \tau\right)
-2n\right)  \frac{\partial \delta \left(  \tau\right)  }{\partial\tau}\right)
}{\partial n}=\allowbreak\frac{1}{3a}\frac{n}{ \delta \left(  \tau\right)
}\frac{\frac{\partial \delta \left(  \tau\right)  }{\partial\tau}}{\left(
n- \delta \left(  \tau\right)  \right)  ^{3}}\left(  n^{2}-3n \delta \left(
\tau\right)  +3\left(   \delta \left(  \tau\right)  \right)  ^{2}\right).
\allowbreak$\bigskip

Next we go to ($ \eqref{mi-mi2}$)

ln$W=-\int\frac{\partial\theta}{\partial n}\frac{\int\frac{\partial T\left(
n,\varepsilon\tau\right)  }{\partial\tau}dn}{a\frac{\partial T\left(
n,\varepsilon\tau\right)  }{\partial n}}d\tau=$

$-\int\left(  \frac{1}{3a}%
\frac{n}{ \delta \left(  \tau\right)  }\frac{\frac{\partial \delta \left(
\tau\right)  }{\partial\tau}}{\left(  n- \delta \left(  \tau\right)  \right)
^{3}}\left(  n^{2}-3n \delta \left(  \tau\right)  +3\left(   \delta \left(
\tau\right)  \right)  ^{2}\right)  \right)  \frac{\left(  \frac{1}{3}%
\frac{n^{2}}{\left(   \delta \left(  \tau\right)  \right)  ^{3}}\left(
T_{w}-T_{\infty}\right)  \left(  3 \delta \left(  \tau\right)  -2n\right)
\frac{\partial \delta \left(  \tau\right)  }{\partial\tau}\right)  }{a\left(
-\frac{1}{\left(   \delta \left(  \tau\right)  \right)  ^{2}}\left(
T_{w}-T_{\infty}\right)  \left(  2 \delta \left(  \tau\right)  -2n\right)
\right)  }d\tau.$

$   \frac{1}{3a}\frac{n}{ \delta \left(  \tau\right)  }\frac
{\frac{\partial \delta \left(  \tau\right)  }{\partial\tau}}{\left(  n- \delta
\left(  \tau\right)  \right)  ^{3}}\left(  n^{2}-3n \delta \left(
\tau\right)  +3\left(   \delta \left(  \tau\right)  \right)  ^{2}\right)
  \frac{\left(  \frac{1}{3}\frac{n^{2}}{\left(   \delta \left(
\tau\right)  \right)  ^{3}}\left(  T_{w}-T_{\infty}\right)  \left(  3 \delta
\left(  \tau\right)  -2n\right)  \frac{\partial \delta \left(  \tau\right)
}{\partial\tau}\right)  }{a\left(  -\frac{1}{\left(   \delta \left(
\tau\right)  \right)  ^{2}}\left(  T_{w}-T_{\infty}\right)  \left(  2 \delta
\left(  \tau\right)  -2n\right)  \right)  }= -\frac{1}{18}%
n^{3}\left(  2n-3 \delta \left(  \tau\right)  \right)  \frac{\partial \delta
\left(  \tau\right)  }{\partial\tau}^{2}\frac{-3n \delta \left(  \tau\right)
+3\left(   \delta \left(  \tau\right)  \right)  ^{2}+n^{2}}{a^{2}\left(
 \delta \left(  \tau\right)  \right)  ^{2}\left(  n- \delta \left(
\tau\right)  \right)  ^{4}},\allowbreak$

ln$W=\frac{n^{3}}{18a^{2}}\int\left(  \left(  2n-3 \delta \left(  \tau\right)
\right)  \left(  \frac{\partial \delta \left(  \tau\right)  }{\partial\tau
}\right)  ^{2}\frac{-3n \delta \left(  \tau\right)  +3\left(   \delta \left(
\tau\right)  \right)  ^{2}+n^{2}}{\left(   \delta \left(  \tau\right)
\right)  ^{2}\left(  n- \delta \left(  \tau\right)  \right)  ^{4}}\right)
d\tau.$

The geometry of the flow model is defined by

\begin{tabular}
[c]{l}%
$dy=\frac{1}{\mu W}\left(  \sin\left(  \theta\right)  d\tau-\mu\cos\left(
\theta\right)  dn\right)  =\frac{\sin\left(  \theta\right)  }{\mu W}%
d\tau-\frac{\cos\left(  \theta\right)  }{W}dn,\allowbreak$\\
$dx=\frac{1}{\mu W}\left(  \cos\left(  \theta\right)  d\tau+\mu\sin
(\theta)dn\right)  =\frac{\cos\left(  \theta\right)  }{\mu W}d\tau+\frac
{\sin\left(  \theta\right)  }{W}dn.\allowbreak$%
\end{tabular}

where

 $\mu=\frac{a\frac{\partial T\left(  n,\tau\right)  }{\partial n}}%
{\int\frac{\partial T\left(  n,\tau\right)  }{\partial\tau}dn}=\frac{-\frac
{a}{\left(   \delta \left(  \tau\right)  \right)  ^{2}}\left(  T_{w}%
-T_{\infty}\right)  \left(  2 \delta \left(  \tau\right)  -2n\right)  }%
{\frac{1}{3}\frac{n^{2}}{\left(   \delta \left(  \tau\right)  \right)  ^{3}%
}\left(  T_{w}-T_{\infty}\right)  \left(  3 \delta \left(  \tau\right)
-2n\right)  \frac{\partial \delta \left(  \tau\right)  }{\partial\tau}%
}=\allowbreak-3\frac{a}{n^{2}} \delta \left(  \tau\right)  \frac{2 \delta
\left(  \tau\right)  -2n}{\left(  3 \delta \left(  \tau\right)  -2n\right)
\frac{\partial \delta \left(  \tau\right)  }{\partial\tau}},\allowbreak$

$dy=\frac{\sin\left(  \theta\right)  }{\mu W}d\tau-\frac{\cos\left(
\theta\right)  }{W}dn,\allowbreak$

$dx=\frac{\cos\left(  \theta\right)  }{\mu W}d\tau+\frac{\sin\left(
\theta\right)  }{W}dn$.

\section{Conclusions}

 We conclude that the mathematical modelling of the convective heat transfer may be realazied in terms of natural for a flow coordinate system via streamline definition. 
We consider this partial problem that link temperature anf velocty fields  as a verification of the proposed approach including the mathematical aspects of the model.

\end{document}